\documentclass[journal,10pt]{IEEEtran}

\usepackage[cmex10]{amsmath}
\usepackage{mathrsfs}

\usepackage{colortbl}
\usepackage[dvipsnames]{xcolor}

\usepackage{amsmath}
\usepackage{amssymb}
\usepackage{stfloats}
\usepackage{cite}
\usepackage{graphicx}
\usepackage{subfigure}
\usepackage{bm}
\usepackage{pifont}
\usepackage{algorithmic}
\usepackage[procnumbered,ruled]{algorithm2e}
\newtheorem{theorem}{\it Theorem}

\begin{document}

\title{Spectral Efficiency and Scalability Analysis for Multi-Level Cooperative Cell-Free Massive MIMO Systems}

\author{Jiamin~Li,~\IEEEmembership{Member,~IEEE},
	Xiaoyu~Sun,~\IEEEmembership{Graduate Student Member,~IEEE},
	Pengcheng~Zhu,
	Dongming~Wang,~\IEEEmembership{Member,~IEEE},
	and~Xiaohu~You,~\IEEEmembership{Fellow,~IEEE}\\

\thanks{This work was supported in part by the National Key R\&D Program of China under Grant 2021YFB2900300, and by the National Natural Science Foundation of China (NSFC) under Grants 61971127, 61871465, 61871122. \emph{(Corresponding author: Pengcheng Zhu.)}}      
\thanks{The authors are with National Mobile Communications Research Laboratory, Southeast University, Nanjing, {\rm 210096}, China (email: \{jiaminli, 220210888, p.zhu, wangdm, xhyu\}@seu.edu.cn). J. Li, D. Wang, and X. You are also with Purple Mountain Laboratories, Nanjing {\rm 211111}, China.}

}



\maketitle

\begin{abstract}


This paper proposes a multi-level cooperative architecture to balance the spectral efficiency and scalability of cell-free massive multiple-input multiple-output (MIMO) systems.
In the proposed architecture, spatial expansion units (SEUs) are introduced to avoid a large amount of computation at the access points (APs) and increase the degree of cooperation among APs.
We first derive the closed-form expressions of the uplink user achievable rates under multi-level cooperative architecture with maximal ratio combination (MRC) and zero-forcing (ZF) receivers.
The accuracy of the closed-form expressions is verified.
Moreover, numerical results have demonstrated that the proposed multi-level cooperative architecture achieves a better trade-off between spectral efficiency and scalability than other forms of cell-free massive MIMO architectures.


\end{abstract}


\IEEEpeerreviewmaketitle

\section{Introduction}
Cell-free massive multiple-input multiple-output (MIMO) has been proposed as a promising technology for beyond fifth-generation (5G) networks\cite{9586055}. 
In original cell-free massive MIMO systems, all access points (APs) are connected to a central processing unit (CPU) which operates all APs as a massive MIMO network with no cell boundaries to serve all user equipment (UEs) by coherent transmission and reception.
Because of overcoming inter-cell interference caused by cellular topology, cell-free massive MIMO has high spectral efficiency, energy efficiency and coverage probability\cite{9079911}.

In original cell-free massive MIMO systems, each AP needs to detect/precode data for all UEs in uplink/downlink and CPU is responsible for processing the signals of all UEs\cite{9064545}.
As the number of UEs increases, the computational complexity and fronthaul capacity required for each AP will grow linearly (or faster), so the original cell-free massive MIMO is not scalable \cite{9174860}.
Some works have analyzed the scalability of the original cell-free massive MIMO systems\cite{8000355}.
Their core idea is that using all APs to serve each UE is impractical because each UE is only close to a subset of APs.
They form the user-centric AP clusters so that each AP only needs to process signals from a subset of the UEs, thus computational complexity is reduced.
A fully distributed and scalable user-centric architecture for cell-free massive MIMO was proposed in \cite{8761828}, improving system scalability by grouping APs in user-centric clusters. 
Based on \cite{8761828}, \cite{9443536} studied a user association procedure to maximize the sum-rate of UEs in the system with the maximum-ratio-combining at APs.
Considering the computational complexity and fronthaul requirements, \cite{9064545} established a new framework for scalable cell-free massive MIMO by exploiting the dynamic cooperative clustering concept. Scalable algorithms for initial access, pilot allocation, cluster formation, precoding and combining were also provided. 
However, the user-centric network architecture requires APs with strong computing power, and the deployment cost of APs is huge.
Furthermore, the cooperation between APs is limited, which leads to a severe performance penalty in the user-centric network architecture compared to the original cell-free massive MIMO\cite{9650567}.

Inspired and motivated by the above discussion, we propose a multi-level cooperative architecture with inter-AP collaboration and inter-spatial expansion unit (SEU) collaboration to balance the spectral efficiency and scalability of cell-free massive MIMO systems. The main contributions of this paper are as follows:
\begin{itemize}
	\item By introducing a small amount of SEUs between APs and the CPU, the computing process is innovatively transferred from APs to SEUs, which makes APs only act as relay stations to reduce system deployment and operation costs. Moreover, the multi-level collaboration of APs and SEUs improves the degree of cooperation between APs and promotes the spectral efficiency. 	
		
	\item We derive the closed-form expressions of the uplink user achievable rates under multi-level cooperative architecture with maximal ratio combination (MRC) and zero-forcing (ZF) receivers at SEUs, and verify the accuracy of the closed expressions. 
	
	\item The proposed system architecture achieves a good trade-off between spectral efficiency and scalability. Numerical results validate it by comparisons with the original cell-free massive MIMO and fully distributed operational scalable massive MIMO systems.
	
\end{itemize}

$Notation$: Boldface letters denote  vectors (lower case) or matrices (upper case). A $N$-dimensional identity matrix is denoted by $\textbf{I}_N$. ${\bf{h}} \sim {\cal C}{\cal N}\left( {0,{\sigma ^2}{{\bf{I}}_N}} \right)$ denotes vector ${\bf{h}}$ satisfies circularly symmetric complex Gaussian distribution with mean zero and covariance matrix ${{\sigma ^2}{{\bf{I}}_N}}$. $(\cdot)^{\text{T}}$ and $(\cdot)^{\text{H}}$ represent the transpose operator and conjugate transpose operator respectively. $|\cdot|$ represents the determinant of a matrix or the size of a set, ${\left\|  \cdot  \right\|^2}$ represents the 2-norm of a vector, and $\mathbb{E}\left[\cdot\right]$ denotes the expectation.


\section{System Model}

\subsection{System Configuration and Virtual Cluster Modeling}
We propose a multi-level cooperative architecture to balance the spectral efficiency and scalability of cell-free massive MIMO systems. 
As illustrated in Fig. 1, there are $M$ APs, $L$ SEUs and a CPU in the system to serve  $K$ single antenna UEs.
Each AP is equipped with $N$ antennas and connected to one of the SEUs through backhaul links. The number of APs connected to the $l$-th SEU is ${ M_{l}}$, where $l \in \{ 1, \cdots ,L\} $.

The introduction of SEUs reduces the computing process of APs and the CPU, but it still cannot satisfy scalability when $K$ approaches infinity. For the $k$-th UE, it is probable that only some specific APs serve it. These APs may be distributed in the coverage of multiple SEUs and constitute a virtual cluster serving the $k$-th UE. 
On the one hand, such a setting prevents the APs with poor channel conditions from serving the $k$-th UE and ensures scalability by allowing each AP to serve a limited number of UEs.
On the other hand, AP virtual service clusters subordinated to multiple SEUs can indraught the multi-level cooperation.
To illustrate the multi-level cooperative architecture, we take the third UE marked in Fig. 1 as an example.
For this UE, its virtual service cluster is composed of 4 APs in the shaded part. APs under the same SEU can form a small network that collaborate to serve UEs within the network service scope by a predefined combining technique in SEUs. 
The two SEUs associated with the virtual service cluster of the third UE receive the signal from the connected APs in the virtual cluster and obtain the estimated values respectively, then transmit the estimated values to the CPU to calculate the final value.

\begin{figure}
	\centering
	\includegraphics[scale=0.42]{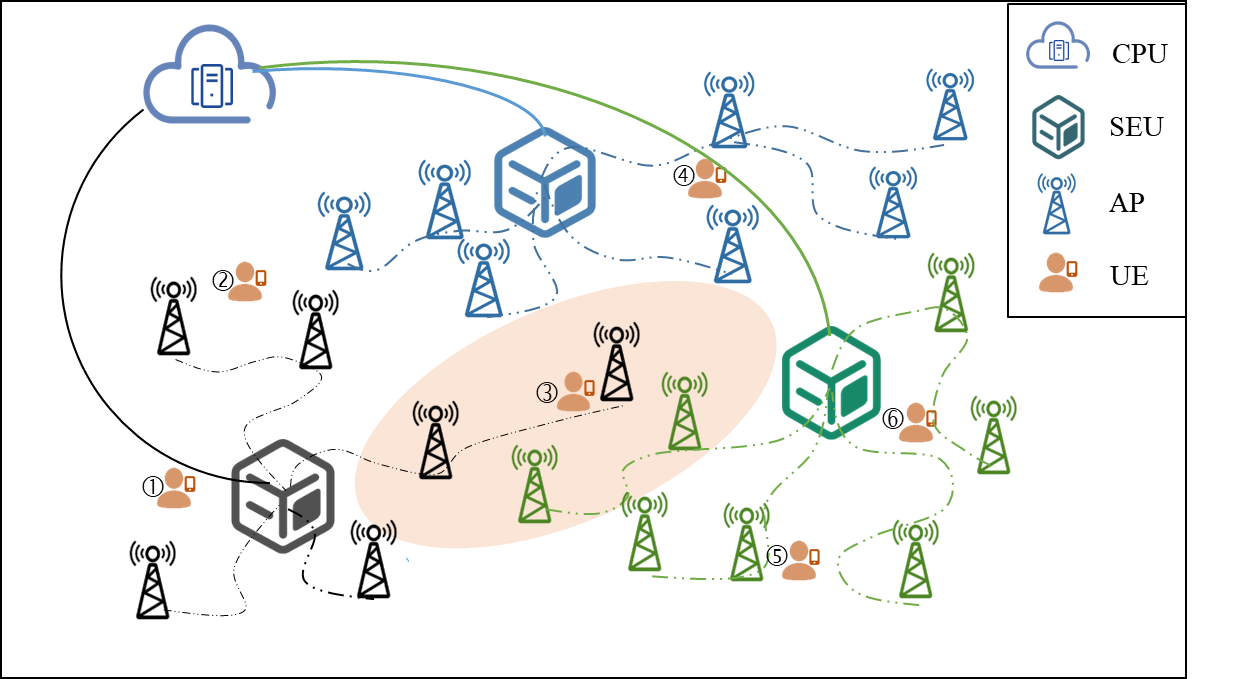}
	\caption{Multi-level cooperative cell-free massive MIMO systems.}
\end{figure}
For notational convenience, we define a AP selection service matrix ${\bf{\Omega}} \in {\mathbb{C}^{{M} \times {K}}}$, the ($i$,$j$)-th element of $\bf{\Omega}$ is 1 if the $i$-th AP is allowed to serve the $j$-th UE and 0 otherwise. 
Such an operation will reduce the number of APs serving each UE, which will inevitably lead to a decline in service quality. Therefore, some generation settings for $\bf{\Omega}$ are designed to minimize the performance loss. 
CPU is set to select the APs whose channel gains are higher than a certain threshold to serve each UE.
According to $\bf{\Omega}$ from CPU, each SEU only need to store the selected service matrix
${{\bf{\Omega}}_l}= \left[ {{{\bm{\omega}}_{l,1}}, \cdots ,{{\bm{\omega}}_{l,K}}} \right] \in \mathbb{C}^{{ M_l} \times {K}} $ 
for subordinate APs, where 
${{\bm{\omega}}_{l,k}} = \left[ {{\upsilon_{l,1,k}}, \cdots ,{\upsilon_{l,{M_l},k}}} \right]^\text{T}$. ${\upsilon_{l,m,k}}$ is 1 to indicate the $m$-th AP of the $l$-th SEU serves the $k$-th UE and 0 otherwise.

The channel vector from the $k$-th UE to all APs is denoted as
\begin{equation}
	{{\bf{g}}_k} = \left({ \bf{\Lambda}} _k^{{1 \mathord{\left/
				{\vphantom {1 2}} \right.
				\kern-\nulldelimiterspace} 2}} \otimes {\bf{I}}_N \right){{\bf{h}}_k},
\end{equation}
where ${\bm{\Lambda} _k} = {\rm{diag}}\left( {{{ {{\lambda _{1,k}},{\lambda _{2,k}}, \cdots ,{\lambda _{M,k}}} }}} \right) $ 
is the covariance matrix, wherein ${\lambda _{m,k}} \buildrel \Delta \over = cd_{m,k}^{ - \alpha }{s_{m,k}}$ represents the large-scale fading, $\alpha$ is the path loss exponent, ${d_{m,k}}$ is the distance from the $k$-th UE to the $m$-th AP, 
and ${\bf{h}}_k  \sim {\cal C}{\cal N}(0,{{\bf I}_{MN}})$ represents the small-scale fast fading.

\subsection{Channel Estimation}
Assuming that the transmission is over frequency-flat fading channels and the system operates in time-division duplex (TDD) mode. 
The channel state information (CSI) is estimated by the SEUs. 
For the $l$-th SEU, $ {\textbf G}_{l} =\left[ {{\textbf{g}}_{l,1}}, \cdots, {{\textbf{g}}_{l,K}} \right] \in \mathbb{C}^{{M_l}N \times K}$ represents the channel matrix from all subordinate APs of the $l$-th SEU to $K$ UEs, and 
the SEU only need to estimate $ \left[ { {\bm{\omega}}_{l,1}^\text{D} } {{{\textbf{g}}}_{l,1}}, \cdots ,{{\bm{\omega}}_{l,K}^\text{D}} {{{\textbf{g}}}_{l,K}}\right]$,  where ${{\bm{\omega}}_{l,k}^\text{D}} \buildrel \Delta \over=  {\rm{diag}}\left( {\bm{\omega}}_{l,k} \right) \otimes {{\textbf{I}}_N}$.

We consider the case of using $\tau $ mutually orthogonal pilot sequences aided transmission to help the SEUs perform channel estimation,
${\bm{\phi} _1}, \cdots ,{\bm{\phi} _\tau } \in {\mathbb{C}^\tau }$ with ${\left\| {{\bm{\phi} _i }} \right\|^2}{\rm{ = }}1$ are used.
Because of $K \gg \tau$,  pilot sequences are reused by multiple UEs.
The index of the pilot assigned to the $k$-th UE denoted as ${t_k} \in \{ 1, \cdots ,\tau \}$ and the index numbers of the UEs using the ${t_k}$-th pilot are contained in set  ${{\cal P}_{t_k}} = \{i|{t_i} = {t_k} \}$.

The signal matrix $\textbf{Y}_m^{\text{p}}\in {\mathbb{C}^{N \times \tau} }$ received at the $m$-th AP is given as:
\begin{equation}
	\textbf{Y}_m^{\text{p}} = \sum\nolimits_{k=1}^K{\sqrt {{\rho_{\text {p}}}} {\textbf{g}_{m,k}} {\bm{\phi}}_{{t_k}}^\text{H}}+{\textbf{n}_m^{\text{p}}},
\end{equation}
where ${{\rho_{\text {p}}}}$ is the transmit power of pilot sequences, ${\textbf{n}_m^{\text{p}}}$ is the complex additive white Gaussian noise (AWGN) vector.
Through the backhaul links, the $l$-th SEU can estimate the required channel information based on $\textbf{Y}_l^{\text{p}} = [ { { {( {\textbf{Y}_{l,1}^{\text{p}}} )}^\text{T}}, \cdots, {{( {\textbf{Y}_{l,{M_l}}^{\text{p}}})}^\text{T}}}]^\text{T}$, where $\textbf{Y}_{l,m}^{\text{p}}$ represents the pilot signal received by the $m$-th AP connected to the $l$-th SEU.

In order to estimate ${{\bm{\omega}}_{l,k}^\text{D}} {{{\textbf{g}}}_{l,k}}$, the received 
$\textbf{Y}_l^{\text{p}}$ is first projeted onto ${\bm{\phi}}_{{t_k}}$,
\begin{equation}
	{{\mathord{\buildrel{\lower3pt\hbox{$\scriptscriptstyle\smile$}} 
				\over y} }_{l,k}} = \textbf{Y}_l^{\text{p}}{\bm{\phi} _{{t_k}}} = \sqrt {{\rho _{\text {p}}}} {\textbf{g}_{l,k}} + \sum\limits_{i \in {P_{{t_k}}}\backslash \{ k\} } {\sqrt {{\rho _{\text {p}}}} {\textbf{g}_{l,i}}}  + {\textbf{n}_l^{\text{p}}}{\bm{\phi} _{{t_k}}},
\end{equation}
the second term is the interference caused by pilot contamination. The minimum mean-square error (MMSE) estimate of ${{\bm{\omega}}_{l,k}^\text{D}} {{{\textbf{g}}}_{l,k}}$ is expressed as 
\begin{align}
	{{\bar {\textbf{g}}}_{l,k}} &= \mathbb{E}\left\{ {{{ {\bm{\omega}} }_{l,k}^\text{D}}{{\textbf{g}}_{l,k}}{{\mathord{\buildrel{\lower3pt\hbox{$\scriptscriptstyle\smile$}} 
					\over y} }_{l,k}}} \right\}{\left( {\mathbb{E}\left\{ {{{\mathord{\buildrel{\lower3pt\hbox{$\scriptscriptstyle\smile$}} 
							\over y} }_{l,k}}\mathord{\buildrel{\lower3pt\hbox{$\scriptscriptstyle\smile$}} 
					\over y} _{l,k}^\text{H}} \right\}} \right)^{ - 1}}{{\mathord{\buildrel{\lower3pt\hbox{$\scriptscriptstyle\smile$}} 
				\over y} }_{l,k}}\notag \\
	&= \sqrt {{\rho _{\text{p}}}}  \left[\left( { \text{diag} ({{\bm{\omega}} _{l,k}}) {{\bm{\Lambda}} _{l,k}}{\bm{\Psi}} _{{t_k},l}^{ - 1}} \right)  \otimes {{\textbf{I}}_N} \right]{{\mathord{\buildrel{\lower3pt\hbox{$\scriptscriptstyle\smile$}} 
				\over y} }_{l,k}},\notag\\
	&= \sqrt {{\rho _{\text{p}}}} \left[\left( { \text{diag} ({ {\bm{\omega}} _{l,k}}) {{\bm{\Lambda}} _{l,k}}{\bm{\Psi}} _{{t_k},l}^{ - 1/2}} \right)  \otimes {{\textbf{I}}_N}\right]{\hat{{\bf h}}}_{l,k} ,	 	
\end{align}
where ${\bm{\Lambda} _{l,k}} = {\rm{diag}}\left( {{{ {{\lambda _{l,1,k}}, \cdots ,{\lambda _{l,{M_l},k}}} }}} \right) $ , ${\bm{\Psi}} _{{t_k},l} = {\rho _{\text {p}}} \sum\limits_{i \in {P_{{t_k}}} } {  {\bm{\Lambda} _{l,k}} + \sigma_{{\text {p}},l}^2{{\textbf{I}}_{M_l}} }$.

To sum up, the estimated channel obtained by the $l$-th SEU 
is $\bar {\bf G}_{l} = \left[\bar {\bf g}_{l,1}, \cdots,\bar {\bf g}_{l,K} \right]$, and
\begin{align}\label{channel}
	\bar{\textbf{g}}_{l,k} = \left[ \kappa_{l,1,k}\hat{\textbf{h}}_{l,1,k}^{\text{T}} , \cdots , \kappa_{l,{ M_l},k}\hat{\textbf{h}}_{l,{ M_l},k}^{\text{T}} \right]^{\text{T}},
\end{align}
where
\begin{align}
\kappa_{l,m,k}& \buildrel \Delta \over = \upsilon_{l,m,k}{\beta _{l,m,k}} ,\\	
{\beta _{l,m,k}}& \buildrel \Delta \over = \frac{{{\lambda _{l,m,k}}}}{{\sqrt {\sum\limits_{k' \in {{\cal P}_{t_k}}} {{\lambda _{l,m,k'}}}  + 1/{\gamma _{\text {p}}}} }},
\end{align}
${\gamma _{\text {p}}}$ is the training signal-to-noise ratio, ${{\bf{\hat h}}_{l,m,k}} \sim {\cal C}{\cal N}(0,{{\rm {\bf{I}}}_{N}})$ represents the equivalent Rayleigh fading part of the estimated channel.
Therefore, we get
\begin{equation}
\begin{aligned}
{{\bar {\textbf{g}}}_{l,k}} \sim {\cal C}{\cal N} (0,\left[{  {\rm{diag}}({\bm{\omega}}_{l,k}) }{{\textbf{R}}_{l,k}}\right] \otimes {{\textbf{I}}_N}),
\end{aligned}
\end{equation}
where ${{\textbf{R}}_{l,k}} = {\rm{diag}}(\beta _{l,1,k}^2, \cdots ,\beta _{l,{M_l},k}^2)$. Furthermore, from the orthogonal property of the MMSE estimation, $ {\tilde{\bf e}}_{l,k}  = {\bm{\omega}}_{l,k}^\text{D} {{\bf{g}}_{l,k}} - {\bar{\bf g}}_{l,k}$,
where ${\tilde{\bf e}}_{l,k} \sim {\cal C}{\cal N}(0,\left[{\rm{diag}}({\bm{\omega}}_{l,k}) ({\bm{\Lambda} _{l,k}} - {{\bf{R}}_{l,k}})\right] \otimes {{\bf I}_N})$ is the uncorrelated estimation error.

\section{User Ergodic Achievable Uplink Rate}
The first part of this section describes the uplink signal model in the proposed multi-level cooperative cell-free massive MIMO systems. In the second part, we derive the closed-form expressions of the user ergodic achievable uplink rates under multi-level cooperative architecture with MRC and ZF receivers.

\subsection{Uplink Signal Model}
APs receive the signals sent by UEs in each coherence block and forward them directly to the connected SEU, which means APs act as relay stations.
The signal received by the $m$-th AP can be expressed as
\begin{equation} 
	{\bf{y}}_m^{\text{ul}} = \sum\nolimits_{k=1}^K{\sqrt{\rho_k} {\bf{g}}_{m,k} x_k} + {{\bf{n}}_m^\text{ul}},
\end{equation}
where ${{\rho _k}}$ and ${{x_k}}$ represent the uplink transmit power and the uplink transmission data of the $k$-th UE respectively, and ${{\bf{n}}_m^\text{ul}} \sim {\cal C}{\cal N}(0,\sigma _{\text{ul},m}^2{\bf{I}}_{N})$.
Through the backhaul links connecting the subordinate APs, the signal received by the $l$-th SEU is ${\bf{y}}_l^{\text{ul}} = [ ({\bf{y}}_{l,1}^{\text{ul}})^{\rm{T}}, \cdots ,({\bf{y}}_{l,{M_l}}^{\text{ul}})^{\rm{T}} ]^{\rm{T}}$.

During the local data processing of the $l$-th SEU, we selectively detect ${\bf{y}}_l^{\text{ul}}$ according to ${\bm{\Omega}}_l$, and get
\begin{equation}\label{eq11}
{{\hat x}_{l,k}} = {{\bf{v}}_{l,k}^{\rm{H}}} {  {\bm{\omega}}_{l,k}^{\text{D}} } {{\bf{y}}_l^{\text{ul}}}
= {{\bf{w}}_{l,k}^{\rm{H}}} {\bf{y}}_l^{\text{ul}}  .
\end{equation}
Note that this estimate is zero for the $k$-th UE not served by subordinate APs of the $l$-th SEU.
For uplink transmission, this paper pays attention to two linear receivers of practical interest, namely MRC and ZF, which are denoted as ${{\bf{v}}_{l,k}} = {[{{\bf{v}}_{l,1,k}}^{\rm{T}}, \cdots ,{{\bf{v}}_{l,{M_l},k}}^{\rm{T}}]^{\rm{T}}}$. Specifically, ${{\bf{w}}_{l,k}}$ can be presented by 
\begin{equation}
	{{\bf{w}}_{l,k}} = \left\{ {\begin{array}{*{20}{c}}
			{{{\bar{{\bf g}}}_{l,k}}}, &{\text{for}{\kern 1pt} {\kern 1pt} {\kern 1pt} {\rm{MRC}}},\\
			{{{\rm{{\bf{b}}}}_{l,k}}}, &{\text{for}{\kern 1pt} {\kern 1pt} {\kern 1pt} {\rm{ZF}},{\kern 1pt} {\kern 1pt} {\kern 1pt}{\kern 1pt} {\kern 1pt} {\kern 1pt}{\kern 1pt} {\kern 1pt} {\kern 1pt}  }
	\end{array}} \right.
\end{equation}
where ${{\bf{b}}_{l,k}}$ is the $k$-th column of $\bar {\bf G}_{l} {(\bar {\bf G}_{l}^{\rm{H}}\bar {\bf G}_{l})^{ - 1}}$.

Since the APs in the virtual cluster of the $k$-th UE may be connected to different SEUs, each SEU sent ${{{\hat x}}_{l,k}}$ to the CPU to calculate the optimal result as
\begin{equation}\label{eq12}
	{{\hat x}_k} = \sum\nolimits_{l = 1}^L {{\mu _{l,k}}    {{{\hat x}}_{l,k}}} .
\end{equation}
Intuitively, the CPU should assign a larger weight to an SEU with a large  signal-to-interference-plus-noise ratio (SINR), but the interference situation and receive combining scheme also should be taken into consideration.  
We use scalable large-scale
fading decoding (LSFD) proposed in \cite{9064545} for weighting at the CPU.
\begin{equation} 
	{\bm{\mu} _k} = {\rho _k}{\left( {\sum\limits_{i \in { {\cal{S}}_k} } {{\rho _i}\mathbb{E}\left\{ {{{\bf{z}}_{ki}}{\bf{z}}_{ki}^\text{H}} \right\}}  + {{\bf{W}}_k} + {{\bf{D}}_k}} \right)^{ - 1}}  \mathbb{E}\left\{ {{{\bf{z}}_{kk}}} \right\},
\end{equation}
where
\begin{center}
	$\begin{aligned}
		{\bm{\mu} _k} &= \left[ {{{\mu} _{1,k}}, \cdots ,{{\mu} _{L,k}}} \right]^{\rm{T}},\\
		{{\bf{z}}_{ki}} &= {\left[ {{\bf{w}}_{1,k}^{\rm{H}}{{\bf{g}}_{1,i}}, \cdots ,{\bf{w}}_{L,k}^{\rm{H}}{{\bf{g}}_{L,i}}} \right]^{\rm{T}}},\\
		{{\bf{W}}_k} &= \sigma _{\text{ul},l}^2 {\rm{diag}}\left( {\mathbb{E}\left\{ {{{\left\| {{{\bf{w}}_{l,k}}} \right\|}^2}} \right\}, \cdots ,\mathbb{E}\left\{ {{{\left\| {{{\bf{w}}_{L,k}}} \right\|}^2}} \right\}} \right),
	\end{aligned}$
\end{center} 
and ${{\bf{D}}_k}\in \mathbb{R}^{L \times L}$ is the diagonal matrix with the ($l$,$l$)-th element being 1 if APs connected to the $l$-th SEU do not serve the $k$-th UE and 0 otherwise, ${ {\cal{S}}_k}$ denotes the set of UEs whose virtual serving clusters partially overlap with that of the $k$-th UE.

According to \eqref{eq11} and \eqref{eq12}, we can get	
\begin{align}
		{{{\hat x}}_k}&= \sum\limits_{l = 1}^L { { {\mu_{l,k}}{ f}_{kk}{x_k}} }
		+ \sum\limits_{l = 1}^L {{\sum\limits_{i \in {{\cal P}_{t_k}}\backslash \{ k\} } {{\mu_{l,k}} { f}_{ki}{x_{i}}} } }\notag\\
		&+ \sum\limits_{l = 1}^L {{\sum\limits_{u \notin {{\cal P}_{t_k}}} { {\mu_{l,k}} { f}_{ku}{x_{u}}} } }  
		+ \sum\limits_{l = 1}^L { {\sum\limits_{k' = 1}^K {{\mu_{l,k}} {\bf{w}}_{l,k}^{\rm{H}}{ {\tilde{\bf e}}_{l,k'}}{x_{k'}}} } }\notag\\
		&+ \sum\limits_{l = 1}^L {{\mu_{l,k}}{\bf{w}}_{l,k}^{\rm{H}}{{\bf{n}}_l^{\text{ul}}}},\label{model}
\end{align}
where ${ f}_{ki} \buildrel \Delta \over=\sqrt {{\rho _i}} {\bf{w}}_{l,k}^{\rm{H}}{{{\bar{\bf g}}}_{l,i}}$.

Treating the interference including noise as worst-case Gaussian distributed noise, we can obtain the ergodic achievable rate of the $k$-th UE which is given by \cite{8421208}
\begin{align} 
{R_k} &= \mathbb{E}\left[ {{{\log }_2}\left( {1 +{\text{SINR}}_k} \right)} \right],\label{biaodashi1}
\end{align}

where
\begin{align}
{\text{SINR}}_k&= \frac{{{\rho _k}\sum\limits_{l = 1}^L {|{\mu_{l,k}}{\bf{w}}_{l,k}^{\rm{H}}{{\bar {\bf{g}}}_{l,k}}{|^2}} }}{{\mathbb{E}\left\{\left[ \sum\limits_{l = 1}^L {\mu_{l,k}^2} (\Upsilon +\sigma _{{{\text{ul},l}}}^2 ||{{\bf{w}}_{l,k}}|{|^2})\right]|{{\bar {\bf G}}_{l}}\right\}}},\label{biaodashi2}\notag\\
{\Upsilon} &= \sum\limits_{i \in {{\cal P}_{t_k}}\backslash \{ k\} } { {\rho _{i}}|{\bf{w}}_{l,k}^{\rm{H}}{{\bar {\bf{g}}}_{l,i}}{|^2}} +{\sum\limits_{u \notin {{\cal P}_{t_k}} } { {\rho _{u}}|{\bf{w}}_{l,k}^{\rm{H}}{{\bar {\bf{g}}}_{l,u}}{|^2}} }\notag\\
&+{\sum\limits_{k' = 1}^K {{\rho _{k'}}|{\bf{w}}_{l,k}^{\rm{H}}{{\tilde {\bf{e}}}_{l,k'}}{|^2}}}. \notag
\end{align}

Based on the equivalent channel model \eqref{biaodashi1}, we carry out spectral efficiency analysis by using achievable rates.

\subsection{Spectral Efficiency Analysis}
The magnitude squared of small-scale fading channel which is assumed to be Rayleigh is Chi-Square distributed. Some lemmas about the projection of isotropic vectors can help give the isotropic approximation.
From lemmas in \cite{8421208}, the channel estimate at the $l$-th SEU can be expressed as
\begin{equation}
{ {\bar{\bf g}}_{l,k}} ^{\rm{H}}{{\bar{\bf g}}_{l,k}} = \sum\limits_{m = 1}^{M_l} {\upsilon _{l,m,k} \beta _{l,m,k}^2{\hat{\bf h}}_{l,m,k}^{\rm{H}} {{\hat{\bf h}}_{l,m,k}}}\sim \Gamma ({\varphi _{l,k}},{\theta _{l,k}}),  
\end{equation}
where
\[\begin{aligned}
{\varphi_{l,k}} &= \frac{{N{{(\sum\nolimits_{m = 1}^{M_l} {\upsilon _{l,m,k}\beta _{l,m,k}^2} )}^2}}}{{\sum\nolimits_{m = 1}^{M_l} ({\upsilon _{l,m,k}\beta _{l,m,k}^2})^2 }},\\
{\theta _{l,k}} &= \frac{{\sum\nolimits_{m = 1}^{M_l} ({\upsilon _{l,m,k}\beta _{l,m,k}^2})^2 }}{{\sum\nolimits_{m = 1}^{M_l} {\upsilon_{l,m,k}\beta _{l,m,k}^2} }}.
\end{aligned}\]

This paper analyzes the spectral efficiency in multi-level cooperative cell-free massive MIMO systems. The theorems provide the closed-form expressions for the uplink achievable rates with MRC and ZF receivers.

\begin{theorem}\label{th1}
The closed-form expression for the uplink achievable rate of the $k$-th UE with MRC receiver in multi-level cooperative cell-free massive MIMO systems is 
\begin{align}
R_k^{\text{mrc}} &= {\log _2}\left(1 +  \frac{{\sum\limits_{l = 1}^L {{\mu_{l,k}^2}\left[N{\rho _k} \left(\sum\limits_{m = 1}^{M_l} {\upsilon _{l,m,k}\beta _{l,m,k}^2}\right)^2 \right]} }}{{\sum\limits_{l = 1}^L {{\mu_{l,k}^2}\left( {{\Upsilon^{\text{mrc}}}+ \sigma _{{{\text{ul},l}}}^2{\varphi} _{l,k}{\theta}_{l,k}} \right)} }} \right),\label{eq13}\\
{\Upsilon^{\text{mrc}}} &= {I_1^{\text{mrc}}}+{I_2^{\text{mrc}}}+{I_3^{\text{mrc}}},
\end{align}
where
\[
\begin{aligned}
	{I_1^{\text{mrc}}} &= {\sum\nolimits_{i \in {{\cal P}_{t_k}}\backslash \{ k\} } { {\rho _i}N(\sum\nolimits_{m = 1}^{M_l} {{\upsilon_{l,m,k}}{\beta _{l,m,k}}{\beta _{l,m,i}}} )} ^2},\\
	{I_2^{\text{mrc}}} &= \sum\nolimits_{u \notin {{\cal P}_{t_k}}} {{\rho _u}\sum\nolimits_{m = 1}^{M_l} {{\upsilon_{l,m,k}}\beta _{l,m,k}^2\beta _{l,m,u}^2} }, \\
	{I_3^{\text{mrc}}} &= \sum\nolimits_{k' = 1}^K {{\rho _{k'}}\sum\nolimits_{m = 1}^{M_l} {{\upsilon_{l,m,k}}\beta _{l,m,k}^2({\lambda _{l,m,k'}} - \beta _{l,m,k'}^2} } ).
\end{aligned}
\]
\end{theorem}
\begin{IEEEproof}
The proof is given in Appendix  $\text{\ref{appendix1}}$.
\end{IEEEproof}

\begin{theorem}\label{th2}
The closed-form expression for the uplink achievable rate of the $k$-th UE with ZF receiver in multi-level cooperative cell-free massive MIMO systems is given by

\begin{align}
R_{k}^{\text{zf}} &= {\log _2}\left( {1 + \frac{{\sum\limits_{l = 1}^L {\mu_{l,k}^2} {\rho _k} \left[{\frac{{ {M_l}N - {K_{l}} + 1}}{{{M_l}N}}{{ \varphi }_{l,k}}{{ \theta }_{l,k}}} \right]}}{{  \sum\limits_{l = 1}^L { {\mu_{l,k}^2} \left(\Upsilon^{\text{zf}} + \sigma _{\text{ul},l}^2 \right)} }}} \right),\label{eq14}\\
\Upsilon^{\text{zf}}&= \sum\nolimits_{k' \in {{\cal U}_{{l}}}} {\frac{{\rho _{k'}}}{{{M_l}N}}}  {{\tilde \varphi }_{l,kk'}}{{\tilde \theta }_{l,kk'}},
\end{align}

where
\[
\begin{aligned}
{\tilde \varphi_{l,kk'}} &= \frac{{N{{(\sum\nolimits_{m = 1}^{M_l} {{{\upsilon_{l,m,k}}}{\eta _{l,m,k'}}} )}^2}}}{{\sum\nolimits_{m = 1}^{M_l} {{\upsilon_{l,m,k}}}{\eta _{l,m,k'}^2} }},\\
{\tilde \theta _{l,kk'}} &= \frac{{\sum\nolimits_{m = 1}^{M_l} \upsilon_{l,m,k}{\eta _{l,m,k'}^2} }}{{\sum\nolimits_{m = 1}^{M_l} \upsilon_{l,m,k}{{\eta _{l,m,k'}}} }},\\
\end{aligned}
\]
${\cal U}{_l} \subset {\rm{ \{ 1,}} \cdots {\rm{,K\}  }}$ indicates the index set of UEs served by APs under the $l$-th SEU,  $K_l$ represents the number of elements in ${\cal U}{_l}$, and ${\eta _{l,m,k}} \buildrel \Delta \over =  {\lambda _{l,m,k}} - \beta _{l,m,k}^2$.

\end{theorem}
\begin{IEEEproof}
The proof is given in Appendix  $\text{\ref{appendix2}}$.
\end{IEEEproof}

\section{Simulation Results}\label{simulation}

In this section, the theoretical results presented in section III are validated. The simulation setup is detailed in the following. A circular area with radius $R$ = 1 km is considered, and all $L = 4$ SEUs, $M = 256$ APs and $K = 32$ UEs are randomly distributed in this area. The minimum access distance from UEs to APs is set as $r_0$ = 30 m. The distance-based path loss model with path loss exponent $\alpha$ = 3.7 is considered. In addition, the uplink transmit power of each UE is ${\rho _{\text{ul},k}}$ = 15 dBm, and the noise signal power is ${\sigma _{\text{ul},l}^2}$ = -84 dBm.

Fig. 2 depicts the simulated results that agree well with the theoretical results when MRC and ZF receivers are adopted by SEUs, which verifies the accuracy of \eqref{eq13} and \eqref{eq14}. 
\begin{figure}
	\centering
	\includegraphics[scale=0.075]{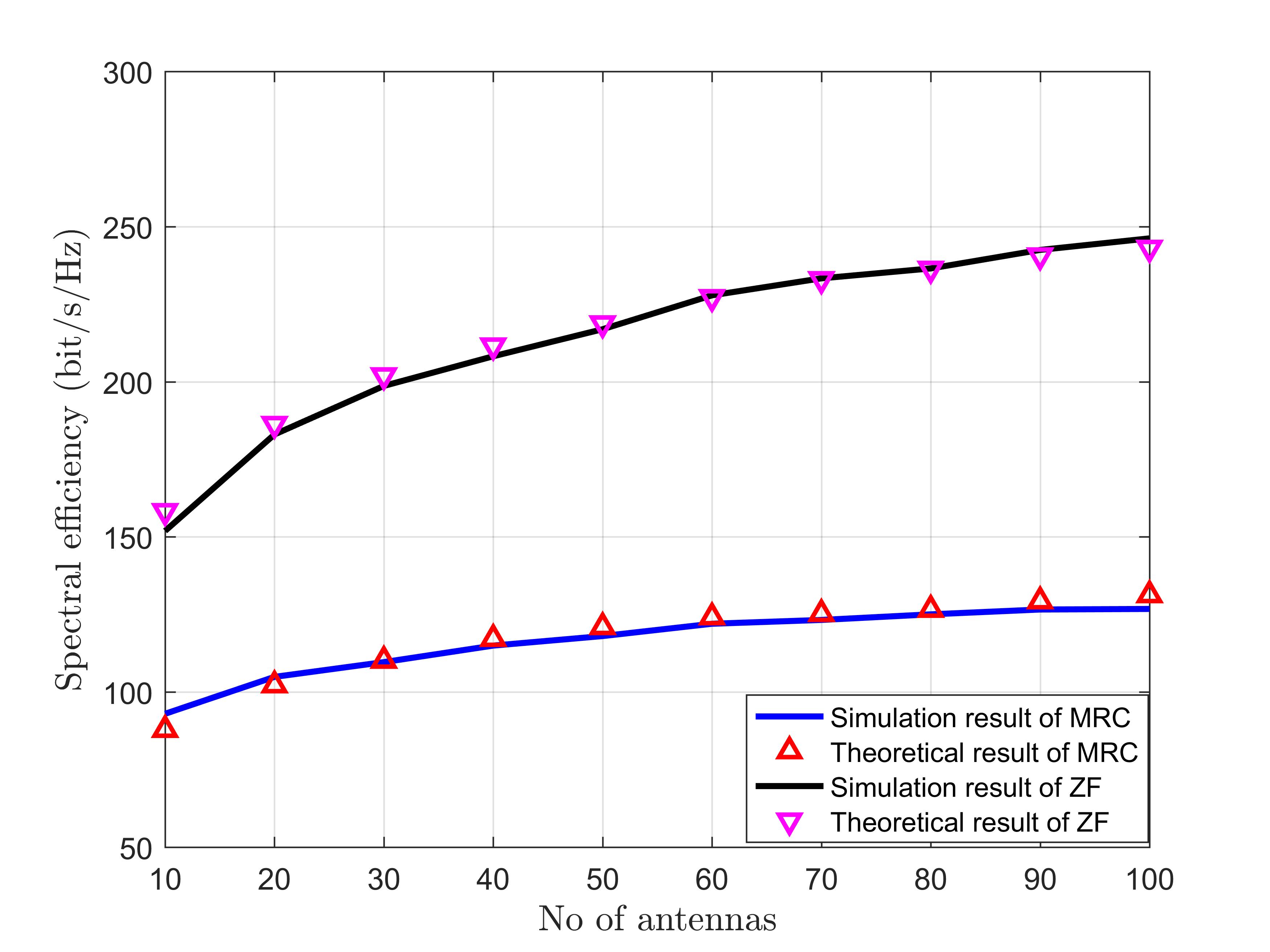}
	\caption{Uplink spectral efficiency against the number of antennas per AP.}
\end{figure}

\begin{figure}
	\centering
	\includegraphics[scale=0.212]{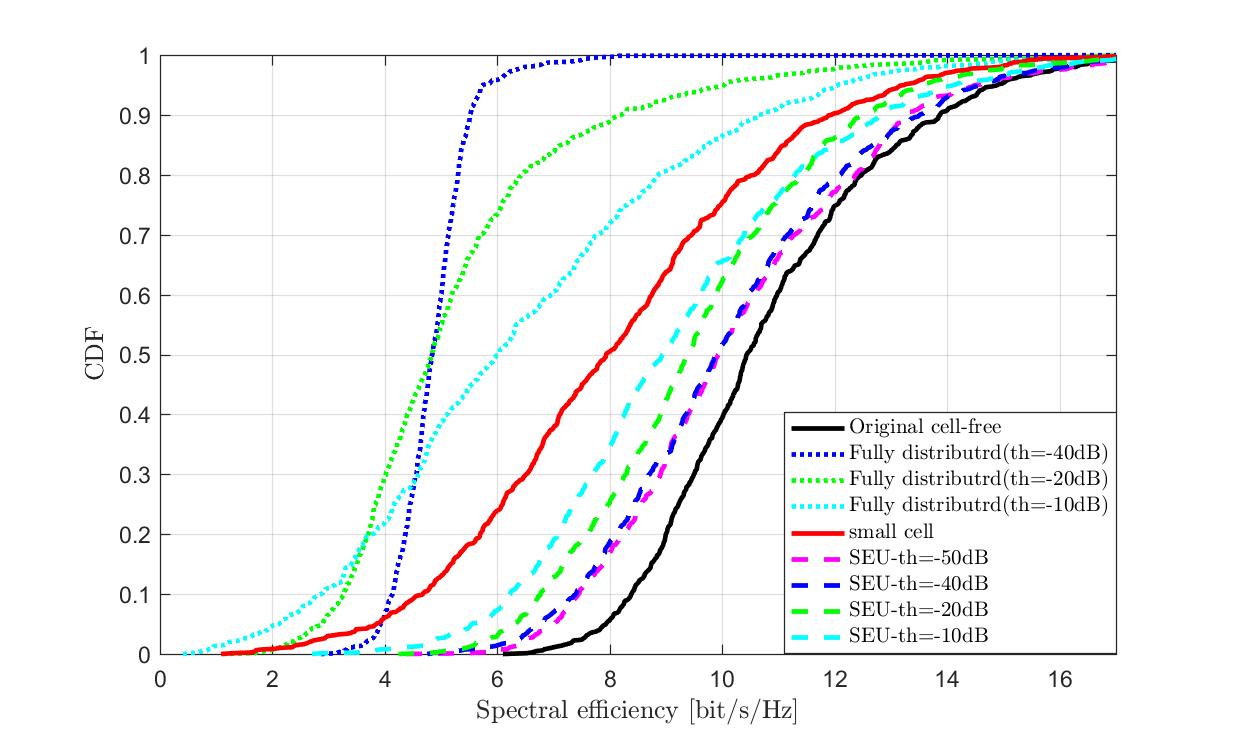}
	\caption{CDFs of uplink spectral efficiency per UE in different systems. }
\end{figure}

Fig. 3 shows the cumulative distribution function (CDF) of the uplink spectral efficiency per UE. 
We compare the multi-level cooperative cell-free massive MIMO systems with original cell-free massive MIMO systems and the fully distributed operation scalable massive MIMO systems.
Fully distributed operation scalable massive MIMO systems also limit the number of UEs served by each AP, but the data is processed individually on each AP and aggregated on the CPU.
Because the SINRs at different APs differ significantly, the performance of the fully distributed operation scalable systems is not as good as that of the systems with multi-level cooperative architecture in the case of the same threshold.
Compared to the original cell-free massive MIMO, the performance loss of the proposed system comes from limiting the number of APs that serve each UE.
However, the performance loss is within the acceptable limit for achieving system scalability. In addition, by adjusting the threshold when the AP selection service matrix is generated, the system performance will approach that of the original cell-free massive MIMO. 


\section{Conclusion}\label{conclu}

In this paper, we propose a multi-level cooperative architecture to obtain a good balance between the spectral efficiency and scalability of cell-free massive MIMO systems.
Firstly, we verify the scalability of the proposed system and derive the closed-form expressions of the uplink user achievable rates with MRC and ZF receivers.
Next, we demonstrate the advantages of multi-level cooperative system architecture by comparison.
It is noted that the proposed architecture has a better trade-off between spectral efficiency and scalability than original cell-free massive MIMO systems and the fully distributed operation scalable massive MIMO systems.

\appendices
\renewcommand{\theequation}{\thesection.\arabic{equation}}
\setcounter{equation}{0}

\section{Proof of Theorem $\text{\ref{th1}}$} \label{appendix1}
To derive the closed-form expression of \eqref{biaodashi1} with MRC receiver, ${\mathbb{E}}[|{\bf{w}}_{l,k}^{\rm{H}}{\bar{\bf{g}}_{l,k}}{|^2}]$ need to be considered. According to ${\bf{w}}_{l,k}^{\rm{H}}{\bar{\bf{g}}_{l,k}} = {\bar{\bf{g}}}_{l,k}^{\rm{H}}{\bar{\bf{g}}_{l,k}}  \sim \Gamma ( {\varphi _{l,k}},{\theta _{l,k}})$ 
\begin{align}
	{\mathbb{E}}\left[{\bf{w}}_{l,k}^{\rm{H}}{\bar{\bf{g}}_{l,k}}\right] &= N\sum\nolimits_{m =1} ^{M_l} {{\upsilon_{l,m,k}}\beta _{l,m,k}^2}, \\
	{\mathbb{E}}\left[|{\bf{w}}_{l,k}^{\rm{H}}{\bar{\bf{g}}_{l,k}}{|^2}\right] &= {\left(N\sum\nolimits_{m =1} ^{M_l} {{\upsilon_{l,m,k}}\beta _{l,m,k}^2} \right)^2}.
\end{align}

Considering ${I_2}$, because the $i$-th UE and the $k$-th UE use the same pilot, we get ${{\hat {\bf{h}}}_{l,m,i}} = {{\hat {\bf{h}}}_{l,m,k}}$ and ${\bf{w}}_{l,k}^{\rm{H}}{{\bar {\bf{g}}}_{l,i}} = \sum\nolimits_{m =1} ^{M_l} {{\upsilon_{l,m,k}}{\beta _{l,m,k}}} {\beta _{l,m,i}}\hat h_{l,m,k}^{\rm{H}}{{\hat h}_{l,m,i}}
\sim {\cal C}{\cal N}( {{\bar \varphi }_{l,ki}},{{\bar \theta }_{l,ki}} )$, where\[\begin{aligned}
{{\bar \varphi }_{l,ki}} &= \frac{{N{{(\sum\nolimits_{m =1} ^{M_l} {{\upsilon_{l,m,k}}{\beta _{l,m,k}}} {\beta _{l,m,i}})}^2}}}{{\sum\nolimits_{m =1} ^{M_l} {({\upsilon_{l,m,k}}{\beta _{l,m,k}}} {\beta _{l,m,i}}{)^2}}},\\
{{\bar \theta }_{l,ki}} &= \frac{{\sum\nolimits_{m =1} ^{M_l} {({\upsilon_{l,m,k}}{\beta _{l,m,k}}} {\beta _{l,m,i}}{)^2}}}{{\sum\nolimits_{m =1} ^{M_l} {{\upsilon_{l,m,k}}{\beta _{l,m,k}}} {\beta _{l,m,i}}}}.
\end{aligned}
\]
Therefore,
\begin{align}
{\mathbb{E}}\left[	{\bf{w}}_{l,k}^{\rm{H}}{{\bar {\bf{g}}}_{l,i}}\right] &= {{\bar \varphi }_{l,ki}}{{\bar \theta }_{l,ki}} ,\\
{\mathbb{E}}\left[|	{\bf{w}}_{l,k}^{\rm{H}}{{\bar {\bf{g}}}_{l,i}}{|^2}\right] &= {\left(N\sum\nolimits_{m =1} ^{M_l} {{\upsilon_{l,m,k}}{\beta _{l,m,k}}} {\beta _{l,m,i}}\right)^2}.
\end{align}

Similarly, ${I_3}$ can be calculated by:
\begin{align}
{\mathbb{E}}[|{\bf{w}}_{l,k}^{\rm{H}}{{{\rm{\tilde {\bf{e}}}}}_{l,k'}}{|^2}] &= {\mathbb{E}}[{\bf{w}}_{l,k}^{\rm{H}}{{{\rm{\tilde  {\bf{e}}}}}_{l,k'}}\tilde  {\bf{e}}_{l,k'}^{\rm{H}}{{\bf{w}}_{l,k}}]\notag\\
& = N\sum\nolimits_{m =1} ^{M_l} {{\upsilon_{l,m,k}}\beta _{l,m,k}^2{\eta _{l,m,k'}}} ,
\end{align}
where ${\eta _{l,m,k}} \buildrel \Delta \over =  {\lambda _{l,m,k}} - \beta _{l,m,k}^2$.  Substituting (A.2), (A.4), (A.5) into \eqref{biaodashi1} can get the closed-from expression  \eqref{eq13}. The proof is completed.

\section{Proof of Theorem $\text{\ref{th2}}$}\label{appendix2}
\setcounter{equation}{0}

For ZF receiver, ${\text{SINR}}_{k}^{\text{zf}}$ in \eqref{biaodashi1} can be expressed as 
\begin{align}
{\text{SINR}}_{k}^{\text{zf}} &=  {\frac{{{\rho _k}\sum\limits_{l = 1}^L {\mu_{l,k}^2} {\mathbb{E}}\left[ {{{\left| {\frac{{{\bf{b}}_{l,k}^{\rm{H}}}}{{||{{\bf{b}}_{l,k}}||}}{{\bar {\bf{g}}}_{l,k}}} \right|}^2}} \right]}}{{\sum\limits_{l = 1}^L { {\mu_{l,k}^2}\left(\sum\limits_{k' \in {{\cal U}_{l}}} {{\rho _{k'}}{\mathbb{E}}\left[ {{{\left| {\frac{{{\bf{b}}_{l,k}^{\rm{H}}}}{{||{{\bf{b}}_{l,k}}||}}{{\tilde {\bf{e}}}_{l,k'}}} \right|}^2}} \right]}   +  \sigma _{{\text{ul}},l}^2 \right) }}}} ,	
\end{align}
${\cal U}{_l} \subset {\rm{ \{ 1,}} \cdots {\rm{,K\}  }}$ indicates the index set of UEs served by APs under the $l$-th SEU.

According to Lemma 3 in \cite{8421208}, we have
${{{\left| {\frac{ {{\bf{b}}_{{\rm{l,k}}}^{\rm{H}}}  }{{||{{\bf{b}}_{l,k}}||}}{{\bar {\bf{g}}}_{l,k}}} \right|}^2}}  \sim \Gamma \left({\frac{{{M_l}N - { K_l} + 1}}{{{ M_l}N}}{ \varphi _{l,k}},{\theta _{l,k}}}\right)$,
where $K_l$ represents the number of elements in ${\cal U}{_l}$. Thus,
\begin{equation}
 {\mathbb{E}}\left[ {{{\left| {\frac{  {{\bf{b}}_{{{l,k}}}^{\rm{H}}}}{{||{{\bf{b}}_{l,k}}||}}{{\bar {\bf{g}}}_{l,k}}} \right|}^2}} \right] = \frac{{{ M_l}N - {K_l} + 1}}{{{ M_l}N}}{ \varphi_{l,k}}{ \theta _{l,k}}.
\end{equation}

Due to the orthogonality property of MMSE estimation, ${{{\tilde {\bf{e}}}_{l,k'}}}$ and ${{{\bf{b}}_{l,k}}}$ are independent. We have
 ${\left| {\frac{ {{\bf{b}}_{{{l,k}}}^{\rm{H}}}}{{||{{\bf{b}}_{l,k}}||}}{{\tilde {\bf{e}}}_{l,k'}}} \right|^2} \sim \Gamma \left({\frac{1}{{{M_l}N}}{\tilde \varphi_{l,kk'}},{\tilde \theta _{l,kk'}}}\right)$, thus
\begin{equation}
	{\mathbb{E}}\left[ {\left| {\frac{{{\bf{b}}_{{{l,k}}}^{\rm{H}}}}{{||{{\bf{b}}_{l,k}}||}}{{\tilde {\bf{e}}}_{l,k'}}} \right|^2} \right] = \frac{1}{{{ M_l}N}}{\tilde  \varphi_{l,kk'}}{\tilde \theta _{l,kk'}}.
\end{equation}	

Substituting (B.2) and (B.3) into \eqref{biaodashi1} yeilds the closed-form expression \eqref{eq14}. The proof is completed.

\ifCLASSOPTIONcaptionsoff
  \newpage
\fi

\bibliographystyle{IEEEtran}
\bibliography{ref}

\end{document}